\documentclass[a4paper,11pt,oneside]{article}

\usepackage[english]{babel}
\usepackage[T1]{fontenc}
\usepackage[utf8]{inputenc}

\usepackage[pdftex,unicode]{hyperref}
\hypersetup{pdftitle=Modeling Price Clustering in High-Frequency Prices}
\hypersetup{pdfauthor=Vladimír Holý and Petra Tomanová}

\usepackage{xcolor}
\definecolor{myred}{rgb}{0.6,0.2,0.3}
\hypersetup{colorlinks=true, linkcolor=myred, anchorcolor=myred, citecolor=myred, filecolor=myred, urlcolor=myred}

\usepackage[margin=58.5pt]{geometry}
\usepackage{amsmath}
\usepackage{amssymb}
\usepackage{amsthm}

\usepackage[group-minimum-digits=3]{siunitx}

\usepackage{graphicx}
\usepackage{float}
\usepackage{hhline}
\usepackage{multirow}
\usepackage{booktabs}
\usepackage{dcolumn}

\usepackage[authoryear]{natbib}

\usepackage{xcolor}

\begin{document}

\begin{center}
{\Large \bfseries Modeling Price Clustering in High-Frequency Prices}
\end{center}

\begin{center}
{\bfseries Vladimír Holý} \\
Prague University of Economics and Business \\
Winston Churchill Square 4, 130 67 Prague 3, Czech Republic \\
\href{mailto:vladimir.holy@vse.cz}{vladimir.holy@vse.cz} \\
Corresponding Author
\end{center}

\begin{center}
{\bfseries Petra Tomanová} \\
Prague University of Economics and Business \\
Winston Churchill Square 4, 130 67 Prague 3, Czech Republic \\
\href{mailto:petra.tomanova@vse.cz}{petra.tomanova@vse.cz}
\end{center}

\begin{center}
{\itshape \today}
\end{center}

\noindent
\textbf{Abstract:}
The price clustering phenomenon manifesting itself as an increased occurrence of specific prices is widely observed and well-documented for various financial instruments and markets. In the literature, however, it is rarely incorporated into price models. We consider that there are several types of agents trading only in specific multiples of the tick size resulting in an increased occurrence of these multiples in prices. For example, stocks on the NYSE and NASDAQ exchanges are traded with precision to one cent but multiples of five cents and ten cents occur much more often in prices. To capture this behavior, we propose a discrete price model based on a mixture of double Poisson distributions with dynamic volatility and dynamic proportions of agent types. The model is estimated by the maximum likelihood method. In an empirical study of DJIA stocks, we find that higher instantaneous volatility leads to weaker price clustering at the ultra-high frequency. This is in sharp contrast with results at low frequencies which show that daily realized volatility has a positive impact on price clustering.
\\

\\

\\

\section{Introduction}
\label{sec:intro}

Over the last two decades, there has been a growing interest in modeling prices at the highest possible frequency which reaches fractions of a second for the most traded assets. The so-called ultra-high-frequency data possess many unique characteristics which need to be accounted for by econometricians. Notably, the prices are irregularly spaced with discrete values. Other empirical properties of high-frequency prices which can be incorporated into models include intraday seasonality, jumps in prices, price reversal and the market microstructure noise. For related models, see, e.g.\ \cite{Russell2005}, \cite{Robert2011}, \cite{Barndorff-Nielsen2012}, \cite{Shephard2017}, \cite{Koopman2017a}, \cite{Koopman2018} and \cite{Buccheri2020}.

We focus on one particular empirical phenomenon observed in high-frequency prices -- price clustering. In general, price clustering refers to an increased occurence of some values in prices. A notable type of price clustering is an increased occurrence of specific multiples of the tick size, i.e.\ the minimum price change. For example, on the NYSE and NASDAQ exchanges, stocks are traded with precision to one cent but multiples of five cents (nickels) and ten cents (dimes) tend to occur much more often in prices. In other words, while one would expect the distribution of the second digit to be uniform, the probability of 0 and 5 is actually higher than 0.1 for each. This behavior can be captured by some agents trading only in multiples of five cents and some only in multiples of ten cents. It is well documented in the literature that this type of price clustering is present in stock markets (see, e.g.\ \citealp{Lien2019}), commodity markets (see, e.g.\ \citealp{Bharati2012}), foreign exchange markets (see, e.g.\ \citealp{Sopranzetti2002}) and cryptocurrency markets (see, e.g.\ \citealp{Urquhart2017}). Moreover, price clustering does not appear only in spot prices but in futures (see, e.g.\ \citealp{Schwartz2004}), options (see, e.g.\ \citealp{ApGwilym2013}) and swaps (see, e.g.\ \citealp{Liu2013a}) as well. From a methodological point of view, almost all papers on price clustering deal only with basic methods and descriptive statistics of the phenomenon. The only paper, to our knowledge, that incorporates price clustering into a price model is the recent theoretical study of \cite{Song2020} which introduced the sticky double exponential jump diffusion process to assess the impact of price clustering on the probability of default.

Our goal is to propose a discrete dynamic model relating price clustering to the distribution of prices and to study the high-frequency behavior of price clustering. We take a fundamentally very different approach than \cite{Song2020} and incorporate the mechanism of an increased occurrence of specific multiples of the tick size directly into the model. This allows us to treat the price clustering phenomenon as dynamic and driven by specified factors rather than given. We also operate within the time series framework rather than the theory of continuous-time stochastic processes. In contrast to the existing literature on price modeling, we do not model log returns or price differences but rather prices themselves. Prices are naturally discrete and positive. When represented as integers, they also exhibit underdispersion, i.e.\ the variance lower than the mean. To accommodate for such features, we utilize the double Poisson distribution of \cite{Efron1986}. It is a less known distribution as noted by \cite{Sellers2017} but was utilized in the context of time series by \cite{Heinen2003}, \cite{Xu2012} and \cite{Bourguignon2019}. Modeling prices directly enables us to incorporate price clustering in the model. Specifically, we consider that prices follow a mixture of several double Poisson distributions with specific supports corresponding to agents trading in different multiples of the tick size. This mixture distribution has a location parameter, a dispersion parameter and parameters determining portions of trader types. In our model, we introduce time variation to all these parameters. We consider the location parameter to be equal to the last observed price resulting in zero expected returns. For the dispersion parameter, we employ dynamics in the fashion of the generalized autoregressive conditional heteroskedasticity (GARCH) model of \cite{Bollerslev1986}. Specifically, we utilize the class of generalized autoregressive score (GAS) models of \cite{Creal2013} and \cite{Harvey2013} which allows to base dynamic models on any underlying distribution. In the high-frequency literature, the GAS framework was utilized by \cite{Koopman2018} for discrete price changes and \cite{Buccheri2020} for log prices. To account for irregularly spaced observations, we include the last trade duration as an explanatory variable similarly to \cite{Engle2000}. Finally, we relate the trader portion parameters to the volatility process and other variables such as the price, the last trade duration and the volume. The resulting observation-driven model is estimated by the maximum likelihood method.

In the empirical study, we analyze 30 Dow Jones Industrial Average (DJIA) stocks in the first half of 2020. We first focus on price clustering from a daily perspective which is a common approach in the price clustering literature. Using a panel regression with fixed effects, we find a positive effect of daily volatility measured by realized kernels of \cite{Barndorff-Nielsen2008} on price clustering. This finding is in line with the results of \cite{ApGwilym1998b, Davis2014, Box2016, Hu2017, Blau2019, Lien2019} among others. Next, we estimate the proposed high-frequency price model and arrive at a different conclusion -- the instantaneous volatility obtained by the model has a negative effect on price clustering. The main message of the empirical study is therefore that the degree of aggregation plays a pivotal role in the relation between price clustering and volatility. While high daily realized volatility correlates with high price clustering, high instantaneous volatility has the opposite effect. The other explanatory variables have the expected effect in both the daily and high-frequency cases -- the volume has a positive effect on price clustering while the price and the last trade duration are insignificant.

The rest of the paper is structured as follows. In Section \ref{sec:lit}, we review the literature dealing with high-frequency price models and price clustering. In Section \ref{sec:model}, we propose the dynamic model accommodating for price clustering based on the double Poisson distribution. In Section \ref{sec:emp}, we use this model to study determinants of price clustering in high-frequency stock prices. We conclude the paper in Section \ref{sec:conclusion}.

\section{Literature Review}
\label{sec:lit}

\subsection{Some High-Frequency Price Models}

In the literature, several models addressing specifics of ultra-high-frequency data have been proposed. One of the key issues is irregularly spaced transactions and discreteness of prices. The seminal study of \cite{Engle1998} proposed the autoregressive conditional duration (ACD) model to capture the autocorrelation structure of trade durations, i.e.\ times between consecutive trades. \cite{Engle2000} combined the ACD model with the GARCH model and jointly modeled prices with trade durations. \cite{Russell2005} again modeled prices jointly with trade durations but addressed discreteness of prices and utilized the multinomial distribution for price changes.

Another approach is to model the price process in continuous time. \cite{Robert2011} considered that the latent efficient price is a continuous Itô semimartingale but is observed at the discrete grid through the mechanism of uncertainty zones. \cite{Barndorff-Nielsen2012} considered the price process to be discrete outright and developed a continuous-time integer-valued Lévy process suitable for ultra-high-frequency data. \cite{Shephard2017} also utilized integer-valued Lévy processes and focused on frequent and quick reversal of prices.

Transaction data at a fixed frequency can also be analyzed as equally spaced time series with missing observations. In this setting, \cite{Koopman2017a} proposed a state space model with dynamic volatility and captured discrete price changes by the Skellam distribution. \cite{Koopman2018} continued with this approach and modeled dependence between discrete stock price changes using a discrete copula. \cite{Buccheri2020} also dealt with multivariate analysis and proposed a model for log prices accommodating for asynchronous trading and the market microstructure noise. The latter two papers utilized the GAS framework.

\subsection{Price Clustering}

The first academic paper on price clustering was written by \cite{Osborne1962}, where the author described the price clustering phenomenon as a pronounced tendency for prices to cluster on whole numbers, halves, quarters, and odd one-eighths in descending preference, like the markings on a ruler. Since then, there have been many studies focusing on this phenomenon -- from \cite{Niederhoffer1965} to very recent papers of \cite{Li2020}, \cite{Song2020}, and \cite{Das2020} -- showing that price clustering is remarkably persistent in time and across various markets.  


\cite{Song2020} pointed out that, however, all studies are entirely focused on empirically examining price clustering in different financial markets. Except for the purely theoretical paper of \cite{Song2020} proposing the sticky double exponential jump-diffusion process to analyze the probability of default for financial variables, the studies related to price clustering are based on basic general methods and do not aim to incorporate the phenomenon into the dynamic price model. 

The prevalent approach to price clustering examination is a linear regression model estimated by ordinary least squares (OLS) method. 
\cite{Ball1985}, \cite{Kandel2001}, and from the recent literature \cite{Urquhart2017}, \cite{Hu2019} and \cite{Li2020}, used the classical regression with dummy variables to estimate frequency of each level of rounding.
The vast of the literature regressed price clustering on explanatory variables such as volatility and trade size, where price clustering is defined as the excess occurrence of multiples of nickles or dimes (see, e.g. \citealp{Schwartz2004} and \citealp{Ikenberry2008} followed by \citealp{Chung2006}, \citealp{Brooks2013}, and \citealp{Hu2017}) or simply their frequency (see, e.g. \citealp{Palao2012} and \citealp{Davis2014}).
However, different definitions of the dependent variable representing price clustering can be found in the literature. \cite{Baig2019} defined the clustering as a sum of round clustering at prices ending by digit 0 and strategic clustering measured as a number of trades which decimals are equal to $01$ or $99$.
\cite{ApGwilym2013} defined the dependent variable as the percentage of price observations at integers, whereas \cite{ApGwilym1998b} estimated the percentage of trades that occur at an odd tick.
\cite{Ahn2005} regressed abnormal even price frequencies in transaction and quote prices on the firm and trading characteristics, and similarly, \cite{Chiao2009} performed the analysis on the limit-order data.
\cite{Cooney2003} estimated cross-sectional regressions of the difference in the percentage of even and odd limit orders on stock price and proxies for investor uncertainty.

Several extensions of the OLS method were employed to overcome certain issues. \cite{Verousis2013} and \cite{Mishra2018} argued that one encounters a simultaneity issue between trade size and price clustering when striving to examine a causal relationship between them. Hence, \cite{Verousis2013} followed by \cite{Mishra2018} used the two-stage least squares (2SLS) method. Moreover, to reflect the endogeneity of quote clustering in the spread model and the endogeneity of the spread in the quote clustering model, \cite{Chung2004, Chung2005} estimated a structural model using three-stage least squares (3SLS) method. \cite{Meng2013} used the 3SLS method to formally examine the hypothesis of a substitution effect between price clustering and size clustering in the CDS market. Finally, \cite{Mbanga2019} estimated robust regressions that eliminate gross outliers to examine the day-of-the-week effect in Bitcoin price clustering. 

Another direction arises from the need to analyze panel data. The prevailing approach is a fixed effects regression. \cite{Das2020} included both firm and time fixed effects, whereas
\cite{Box2016} included fixed effects only for time and report that once they also included firm fixed effects, the results remained unchanged. \citealp{Blau2019} and \cite{Blau2016} included month and year fixed effects respectively, and used robust standard errors that account for clustering across both the cross-sectional observations and time-series observations. On the other hand, \cite{Ohta2006} picked random effects model over the fixed effects model based on the results from the Hausman specification test.

A substantial part of the literature models price clustering as a binary variable. For that case, the straightforward approach is to use the logit or probit model. 
From one of the first papers using logistic regression to analyze price clustering, \cite{Ball1985} modeled three dependent variables taking value 1 if the price is rounded to the whole dollar, half-dollar, or quarter, respectively.
\cite{Christie1994a} estimated logistic regressions that predict the probability of a firm being quoted using odd eighths.
\cite{Aitken1996} employed multivariate logistic regression to model three binary dependent variables that are equal to one if the final digit is 0; 0 or 5; and 0, 2, 4, 6, or 8 (even numbers), respectively. 
\cite{Brown2008} examined the influence of Chinese culture on price clustering by logistic regressions where a binary dependent variable is equal to 1 if the last sale price ends in 4 and 0 if it ends in 8 since many Chinese consider the number 8 as \emph{lucky} while 4 is considered as \emph{unlucky}.
From the literature employing probit models, \cite{Kahn1999} analyzed the propensity to set retail deposit interest rates at integer levels and \cite{Sopranzetti2002} analyzed the propensity for exchange rates to cluster on even digits. 
Moreover, \cite{Capelle-Blancard2007} models a binary dependent variable that is equal to one if the transaction price ends with $00$, whereas \cite{Liu2011} and \cite{Narayan2013} set the variable equal to one if the price ends at either 0 or 5, and 0 otherwise.
\cite{Alexander2007} followed by \cite{Lien2019} used a bivariate probit model to take into account the dependence between price and trade-size clustering. 

Finally, \cite{Blau2019} estimated a vector autoregressive process and examined the impulses of price clustering in response to an exogenous shock to investor sentiment. Besides the classical regression approaches, \cite{Harris1991} and \cite{Hameed1998} analyzed the cross-sectional data by static discrete price model.

To the best of our knowledge, the literature still lacks a discrete dynamic model to study the high-frequency behavior of the price clustering. Thus, in the next section, we propose a novel model which models high-frequency prices directly at the highest possible frequency and allows us to study the main drivers of price clustering such as price, volatility, volume, and trading frequency in the form of trade durations.

\section{Dynamic Price Clustering Model}
\label{sec:model}

\subsection{Double Poisson Distribution}
\label{sec:modelDoublepoiss}

Let us start with the static version of our model for prices. In the first step, we transform the observed prices to have integer values. For example, on the NYSE and NASDAQ exchanges, the prices are recorded with precision to two decimal places and we therefore multiply them by 100 to obtain integer values. The minimum possible change in the transformed prices is 1. Empirically, the transformed prices exhibit strong \emph{underdispersion}, i.e.\ the variance lower than the mean. In our application, the transformed prices are in the order of thousands and tens of thousands while the price changes are in the order of units and tens. We therefore need to base our model on a count distribution allowing for underdispersion. For a review of such distributions, we refer to \cite{Sellers2017}. Although not without its limitations, the double Poisson distribution is the best candidate for our case as the alternative distributions have too many shortcomings. For example, the condensed Poisson distribution is based on only one parameter, the generalized Poisson distribution can handle only limited underdispersion and the gamma count distribution as well as the Conway--Maxwell--Poisson distribution do not have the moments available in a closed form. 

The double Poisson distribution was proposed in \cite{Efron1986} and has a \emph{location} parameter $\mu > 0$ and a \emph{dispersion} parameter $\alpha$. We adopt a slightly different parametrization than \cite{Efron1986} and use the logarithmic transformation for the dispersion parameter making $\alpha$ unrestricted. For $\alpha = 0$, the distribution reduces to the Poisson distribution. Values $\alpha > 0$ result in underdispersion while values $\alpha < 0$ result in overdispersion. Let $Y$ be a random variable and $y$ an observed value. The probability mass function is given by
\begin{equation}
\label{eq:distDoublepoissProb}
\mathrm{P} [Y = y | \mu, \alpha] = \frac{1}{C(\mu,\alpha)} \frac{y^y}{y!} \left( \frac{\mu}{y} \right)^{e^{\alpha} y} e^{\frac{\alpha}{2} + e^{\alpha} y - e^{\alpha} \mu - y},
\end{equation}
where $C(\mu,\alpha)$ is the \emph{normalizing constant} given by
\begin{equation}
\label{eq:distDoublepoissConstant}
C(\mu,\alpha) = \sum_{y=0}^{\infty} \frac{y^y}{y!} \left( \frac{\mu}{y} \right)^{e^{\alpha} y} e^{\frac{\alpha}{2} + e^{\alpha} y - e^{\alpha} \mu - y}.
\end{equation}
The log likelihood for observation $y$ is then given by
\begin{equation}
\label{eq:distDoublepoissLik}
\ell(y; \mu,\alpha) = - \ln \left( C(\mu, \alpha) \right) + y \ln(y) - \ln(y!) + e^{\alpha} y \ln \left( \frac{\mu}{y} \right) + \frac{\alpha}{2} + e^{\alpha} y - e^{\alpha} \mu - y.
\end{equation}
Unfortunately, the normalizing constant is not available in a closed form. However, as \cite{Efron1986} shows, it is very close to 1 (at least for some combinations of $\mu$ and $\alpha$) and can be approximated by
\begin{equation}
\label{eq:distDoublepoissApproxEfron}
C(\mu,\alpha) \simeq 1 + \frac{1 - e^{\alpha}}{12 e^{\alpha} \mu} \left( 1 + \frac{1}{e^{\alpha} \mu} \right).
\end{equation}
\cite{Zou2013} notes that approximation \eqref{eq:distDoublepoissApproxEfron} is not very accurate for low values of the mean and suggest approximating the normalizing constant alternatively by cutting off the infinite sum, i.e.
\begin{equation}
\label{eq:distDoublepoissApproxZou}
C(\mu,\alpha) \simeq \sum_{y=0}^{m} \frac{y^y}{y!} \left( \frac{\mu}{y} \right)^{e^{\alpha} y} e^{\frac{\alpha}{2} + e^{\alpha} y - e^{\alpha} \mu - y},
\end{equation}
where $m$ should be at least twice as large as the sample mean. In our case of high mean, approximation \eqref{eq:distDoublepoissApproxEfron} is sufficient while approximation \eqref{eq:distDoublepoissApproxZou} would be computationally very demanding and we therefore resort to the former one. The expected value and variance can be approximated by
\begin{equation}
\label{eq:distDoublepoissMoments}
\mathrm{E}[Y] \simeq \mu, \qquad \mathrm{var}[Y] \simeq \mu e^{-\alpha}. \\
\end{equation}
The score can be approximated by
\begin{equation}
\label{eq:distDoublepoissScore}
\nabla (y; \mu, \alpha) \simeq \left( \frac{e^{\alpha}}{\mu} (y - \mu),  e^{\alpha} \left( y \ln(\mu) - \mu - y \ln(y) + y \right) + \frac{1}{2} \right)'.
\end{equation}
The Fisher information can be approximated by
\begin{equation}
\label{eq:distDoublepoissFisher}
\mathcal{I} (\mu, \alpha) \simeq 
\begin{pmatrix}
\frac{e^{\alpha}}{\mu} & 0 \\
0 & \frac{1}{2} \\
\end{pmatrix}.
\end{equation}

\subsection{Mixture Distribution for Price Clustering}
\label{sec:modelMixture}

\begin{figure}
\begin{center}
\includegraphics[width=15cm]{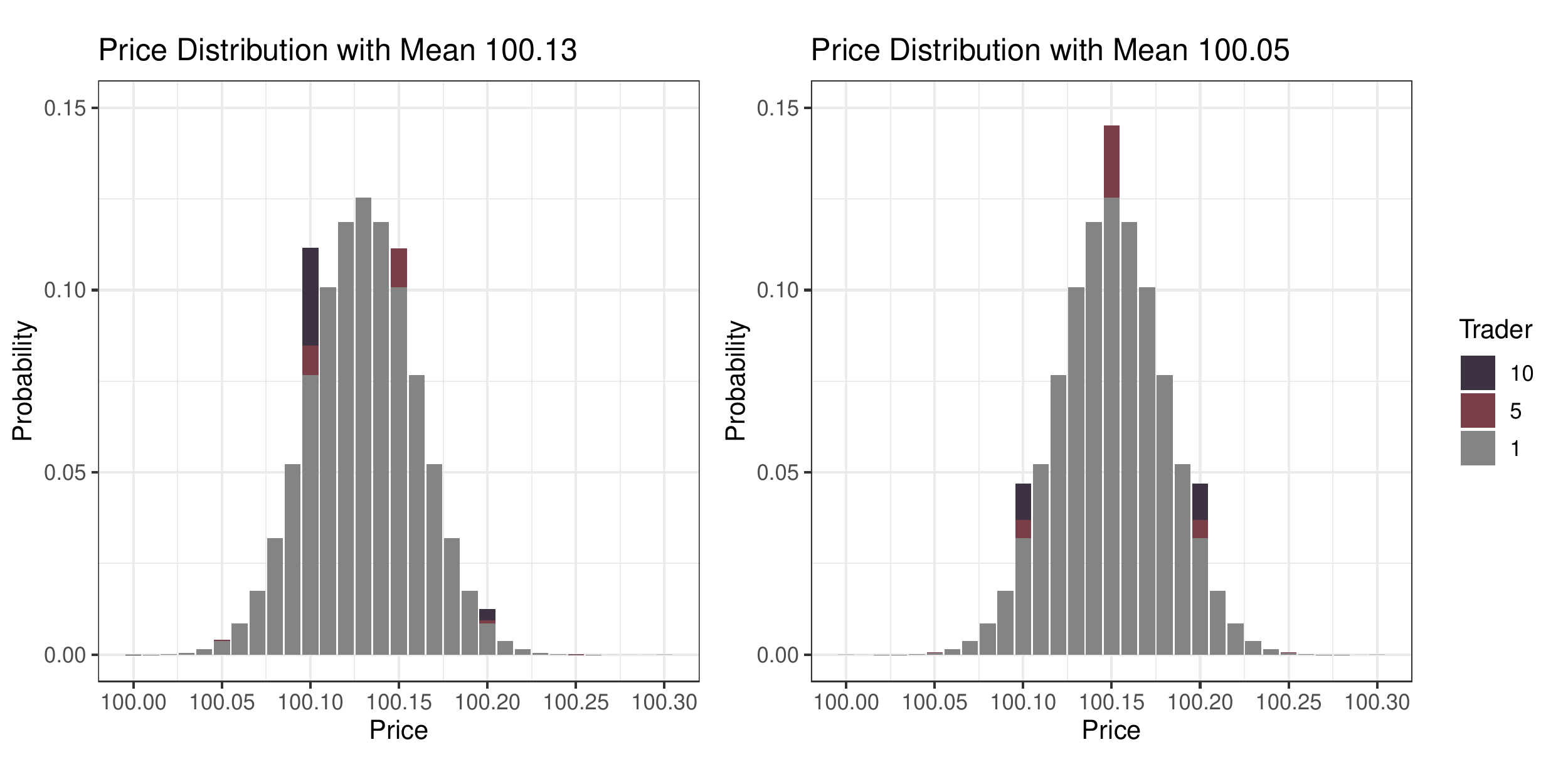} 
\caption{Illustration of the probability mass function for the mixture double Poisson distribution with parameters $\mu = \num{10013}$ (left plot), $\mu = \num{10005}$ (right plot), $\alpha=7$, $\varphi_{1}=0.95$, $\varphi_{5}=0.02$ and $\varphi_{10}=0.03$. The prices are reported in the original form with two decimal places.}
\label{fig:example}
\end{center}
\end{figure}

Next, we propose a mixture of several double Poisson distributions corresponding to trading in different multiples of tick sizes accommodating for price clustering. We consider that there are three types of traders -- one who can trade in cents, one who can trade only in multiples of 5 cents and one who can trade only in multiples of 10 cents. In Appendix \ref{app:mixture}, we treat a more general case with any number of trader types and tick size multiples. The distribution of prices corresponding to each trader type is based on the double Poisson distribution modified to have support consisting only of multiples of $k \in \{ 1, 5, 10 \}$ while keeping the expected value $\mathrm{E}[Y] \simeq \mu$ and the variance $\mathrm{var}[Y] \simeq \mu e^{-\alpha}$ regardless of $k$. For a detailed derivation of the distribution, see Appendix \ref{app:mixture}. The distribution of prices for trader type $k \in \{ 1, 5, 10 \}$ is given by
\begin{equation}
\label{eq:distMixtureTraderProb}
\mathrm{P} \left[ Y^{[k]} = y \middle| \mu, \alpha \right] = \mathbb{I} \left\{ k \mid y \right\} \mathrm{P} \left[ Z^{[k]} = \frac{y}{k} \middle| \mu, \alpha \right], \qquad Z^{[k]} \sim \mathrm{DP} \left( \frac{\mu}{k}, \alpha + \ln(k) \right),
\end{equation}
where $\mathrm{DP}$ denotes the double Poisson distribution and $\mathbb{I} \left\{ k \mid y \right\}$ is equal to 1 if $y$ is divisible by $k$ and 0 otherwise. Note that for $k=1$, it is the standard double Poisson distribution. Finally, the distribution of all prices is the mixture
\begin{equation}
\label{eq:distMixtureProb}
\mathrm{P} \left[ Y = y \middle| \mu, \alpha, \varphi_{1}, \varphi_{5}, \varphi_{10} \right] = \sum_{k \in \{ 1, 5, 10 \}} \varphi_{k} \mathrm{P} \left[ Y^{[k]} = y \middle| \mu, \alpha \right],
\end{equation}
where the parameter space is restricted by $\mu > 0$, $\varphi_{1} \geq 0$, $\varphi_{5} \geq 0$, $\varphi_{10} \geq 0$ and $\varphi_{1} + \varphi_{5} + \varphi_{10} = 1$. Parameters $\varphi_k$, $k \in \{ 1, 5, 10 \}$, are the portions of trader types and parameters $\mu$ with $\alpha$ have the same interpretation as in the double Poisson distribution. The log likelihood for observation $y$ is given by
\begin{equation}
\label{eq:distMixtureLik}
\begin{aligned}
\ell \left( y; \mu,\alpha, \varphi_{1}, \varphi_{5}, \varphi_{10} \right) &= e^{\alpha} y \ln \left( \frac{\mu}{y} \right) + \frac{\alpha}{2} + e^{\alpha} y - e^{\alpha} \mu \\
& \qquad + \ln \left( \sum_{k \in \{ 1, 5, 10 \}} \varphi_{k} \mathbb{I} \left\{ k | y \right\} \frac{ \sqrt{k} }{ C \left( \frac{\mu}{k}, k \alpha \right) } \frac{ \left( \frac{y}{k} \right)^{\frac{y}{k}} }{ \left( \frac{y}{k} \right) ! } e^{-\frac{y}{k}} \right).
\end{aligned}
\end{equation}
Note that the last logarithm in \eqref{eq:distMixtureLik} is not dependent on parameters $\mu$ and $\alpha$ besides the normalizing constant making the approximation of the score quite simple. Additionaly, parameters $\varphi_{1}$, $\varphi_{5}$ and $\varphi_{10}$ appear only in the last logarithm in \eqref{eq:distMixtureLik} making the approximation of the score for parameters $\mu$ and $\alpha$ independent of parameters $\varphi_{1}$, $\varphi_{5}$ and $\varphi_{10}$. The approximations of the expected value and the variance as well as the score and the Fisher information for the parameters $\mu$ and $\alpha$ of the mixture distribution are therefore the same as for the regular double Poisson distribution presented in \eqref{eq:distDoublepoissMoments}, \eqref{eq:distDoublepoissScore} and \eqref{eq:distDoublepoissFisher} respectively when assuming  $C(\mu / k, k \alpha) = 1$ . Figure \ref{fig:example} illustrates the probability mass function of the mixture distribution.

\subsection{Dynamics of Time-Varying Parameters}
\label{sec:modelDynamics}

Finally, we introduce time variation into parameters $\mu, \alpha, \varphi_{1}, \varphi_{5}, \varphi_{10}$. We denote the random prices as $Y_t \in \mathbb{N}_0$, $t = 1, \ldots, n$ and the observed values as $y_t \in \mathbb{N}_0$, $t = 1, \ldots, n$. We also utilize observed trade durations $z_t \in \mathbb{R}^+$, $t = 1, \ldots, n$ and observed volumes $v_t \in \mathbb{R}^+$, $t = 1, \ldots, n$. We assume that $Y_t$ follow the mixture double Poisson distribution proposed in Section \ref{sec:modelMixture} with time-varying parameters $\mu_t$, $\alpha_t$, $\varphi_{1,t}$, $\varphi_{5,t}$ and $\varphi_{10,t}$. The dynamics of the location parameter $\mu_t$ is given by
\begin{equation}
\label{eq:dynamicsMu}
\mu_{t} = y_{t-1}.
\end{equation}
This means that the expected value of the price is (approximately) equal to the last observed price, i.e.\ the expected value of the return is zero.  This is a common assumption for high-frequency returns (see, e.g.\ \citealp{Koopman2017a}).

For the dynamics of the dispersion parameter $\alpha_t$, we utilize the generalized autoregressive score (GAS) model of \cite{Creal2013}, also known as the dynamic conditional score (DCS) model by \cite{Harvey2013}. The GAS model is an observation-driven model providing a general framework for modeling time-varying parameters for any underlying probability distribution. It captures dynamics of time-varying parameters by the autoregressive term and the score of the conditional density function. \cite{Blasques2015} investigated information-theoretic optimality properties of the score function and showed that only parameter updates based on the score will always reduce the local Kullback–Leibler divergence between the true conditional density and the model-implied conditional density. \cite{Creal2013} suggested to scale the score based on the Fisher information. As the Fisher information for the parameter $\alpha_t$ is constant in our case, the score is already normalized and we therefore omit the scaling. Using \eqref{eq:distDoublepoissScore} and \eqref{eq:dynamicsMu}, we let the dispersion parameter $\alpha_t$ follow the recursion
\begin{equation}
\label{eq:dynamicsAlpha}
\alpha_{t} = c + b \alpha_{t-1} + a \left( e^{\alpha_{t-1}} \left( y_{t-1} \ln \left( \frac{y_{t-2}}{y_{t-1}} \right) - y_{t-2} + y_{t-1} \right) + \frac{1}{2} \right) + d \ln (z_{t}),
\end{equation}
where $c$ is the constant parameter, $b$ is the autoregressive parameter, $a$ is the score parameter and $d$ is the duration parameter. This volatility dynamics corresponds to the generalized autoregressive conditional heteroskedasticity (GARCH) model of \cite{Bollerslev1986}. Similarly to \cite{Engle2000}, we also include the preceding trade duration $z_{t}$ as an explanatory variable to account for irregularly spaced observations. To prevent extreme values of durations, we use the logarithmic transformation.

The portions of trader types are driven by process
\begin{equation}
\label{eq:dynamicsEta}
\eta_{t} = f \eta_{t-1} + g_1 \ln(\mu_{t}) + g_2 \left( \ln(\mu_{t}) - \alpha_{t} \right) + g_3 \ln(z_{t}) + g_4 \ln(v_{t}),
\end{equation}
where $f$ is the autoregressive parameter, $g_1$ is the parameter for the logarithm of the expected price, $g_2$ is the parameter for the logarithm of the variance of the price process $\mu_{t} e^{-\alpha_{t}}$, $g_3$ is the parameter for the logarithm of the preceding trade duration, and $g_4$ is the parameter for the logarithm of the volume $v_{t}$. The portions of trader types are then standardized as
\begin{equation}
\label{eq:dynamicsPhi}
\varphi_{1,t} = \frac{e^{\eta_{t}}}{e^{\eta_{t}} + h_{5} + h_{10}}, \quad
\varphi_{5,t} = \frac{h_{5}}{e^{\eta_{t}} + h_{5} + h_{10}}, \quad
\varphi_{10,t} = \frac{h_{10}}{e^{\eta_{t}} + h_{5} + h_{10}},
\end{equation}
where $h_{5} \geq 0$ and $h_{10} \geq 0$ are parameters capturing representation of 5 and 10 trader types. The model can be straightforwardly extended to include additional explanatory variables in \eqref{eq:dynamicsAlpha} and \eqref{eq:dynamicsEta}.

\subsection{Maximum Likelihood Estimation}
\label{sec:modelEst}

The proposed model based on the mixture distribution for price clustering \eqref{eq:distMixtureProb} with dynamics given by \eqref{eq:dynamicsMu}, \eqref{eq:dynamicsAlpha} and \eqref{eq:dynamicsPhi} can be straightforwardly estimated by the conditional maximum likelihood method. Let $\theta = (c, b, a, d, f, g_1, g_2, g_3, g_4, h_{5}, h_{10})'$ denote the static vector of all parameters. The parameter vector $\theta$ is then estimated by the conditional maximum likelihood
\begin{equation}
\hat{\theta} \in \arg\max_\theta \sum_{t=1}^n \ell \left( y_t; \mu_t, \alpha_t, \varphi_{1,t}, \varphi_{5,t}, \varphi_{10,t} \right),
\end{equation}
where $\ell(y_t; \mu_t, \alpha_t, \varphi_{1,t}, \varphi_{5,t}, \varphi_{10,t})$ is given by \eqref{eq:distMixtureLik}.

For the numerical optimization in the empirical study, we utilize the PRincipal AXIS algorithm of \cite{Brent1972}. To improve numerical performance, we standardize the explanatory variables to unit mean. We also run the estimation procedure several times with different starting values to avoid local maxima.

\section{Empirical Results}
\label{sec:emp}

\subsection{Data Sample}
\label{sec:empData}

The empirical study is conducted on transaction data extracted from the NYSE TAQ database which contains intraday data for all securities listed on the New York Stock Exchange (NYSE), American Stock Exchange (AMEX), and Nasdaq Stock Market (NASDAQ). We analyze 30 stocks that form the Dow Jones Industrial Average (DJIA) index in June 2020. The extracted data span over six months from January 2 to June 30, 2020, except for Raytheon Technologies (RTX)\footnote{The RTX company results from the merge of the United Technologies Corporation and the Raytheon Company on April 3, 2020.} for which the data are available from April 3, 2020.  

We follow the standard cleaning procedure for the NYSE TAQ dataset described in \cite{Barndorff-Nielsen2009} since data cleaning is an important step of high-frequency data analysis \citep{Hansen2006}. Before the standard data pre-processing is conducted, we delete entries that are identified as preferred or warrants (trades with the non-empty suffix indicator). Then we follow a common data cleaning steps and discard (i) entries outside the main opening hours (9:30 -- 16:00), (ii) entries with the transaction price equal to zero, (iii) entries occurring on a different exchange than it is primarily listed, (iv) entries with corrected trades, (v) entries with abnormal sale condition, (vi) entries for which the price deviated by more than 10 mean absolute deviations from a rolling centered median of 50 observations, and (vi) duplicate entries in terms of the time stamp. In the last step, we remain the entry with mode price instead of the originally suggested median price due to avoiding distortion of the last decimal digit of prices.

The first and last step has a negligible impact on our data and steps ii, iv, and vi have no impact at all. However, the third step causes a large deletion of the data which is, however, in line with \cite{Barndorff-Nielsen2009}. The basic descriptive statistics after data pre-processing are shown in Appendix~\ref{app:descriptive}. Number of observations ranges from \num{216618} (TRV) to \num{3099279} (MSFT). Price clustering in terms of the excess occurrence of multiples of five cents and ten cents in prices ranges from 1.45 \% (KO) to 11.52 \% (BA). First, we analyze the price clustering using a common approach of fixed effects model on daily data in Section \ref{sec:empDaily} to investigate whether the results for our dataset are in line with the existing literature. Then, we estimate the proposed dynamic price model in Section \ref{sec:empHigh}.

\subsection{Analysis Based on Daily Data}
\label{sec:empDaily}

In this section, we investigate the main determinants of price clustering for which pervasive evidence is documented in the literature, namely price, volatility, trading frequency (which we measure in terms of trade durations), and volume. We use a panel regression with fixed effects to take into account the unobserved heterogeneity in both dimensions -- stocks and days. 

Let us define price clustering $p_{i,t}$ as the excess relative frequency of multiples of five cents and ten cents in prices of stock $i$ at day $t$. We model $p_{i,t}$ as
\begin{equation}\label{eq:plm}
p_{i,t} = \gamma_{i} + \delta_{t} + \beta_1 \ln(\overline{\mu}_{i,t}) + \beta_2 \ln(\overline{\alpha}_{i,t}) + \beta_3 \ln(\overline{z}_{i,t}) + \beta_4 \ln(\overline{v}_{i,t}) + \varepsilon_{i,t},
\end{equation} 
where $\gamma_{i}$ is a stock specific effect for the stock $i$, $\delta_{t}$ is a time effect for day $t$, and $\varepsilon_{i,t}$ is the error term. Parameters $\beta_1$ -- $\beta_4$ corresponds to logarithmic explanatory variables, where $\overline{\mu}_{i,t}$ is an average price, $\overline{z}_{i,t}$ is an average duration and $\overline{v}_{i,t}$ is an average volume, where all averages are calculated for each stock $i$ at each day $t$. Daily volatility $\overline{\alpha}_{i,t}$ is estimated by realized kernel estimator of \cite{Barndorff-Nielsen2008}. We use Parzen kernel as suggested by \cite{Barndorff-Nielsen2009}. See \cite{Holy2020c} for a comprehensive overview of quadratic covariation estimators.

\begin{figure}
\begin{center}
\includegraphics[width=15cm]{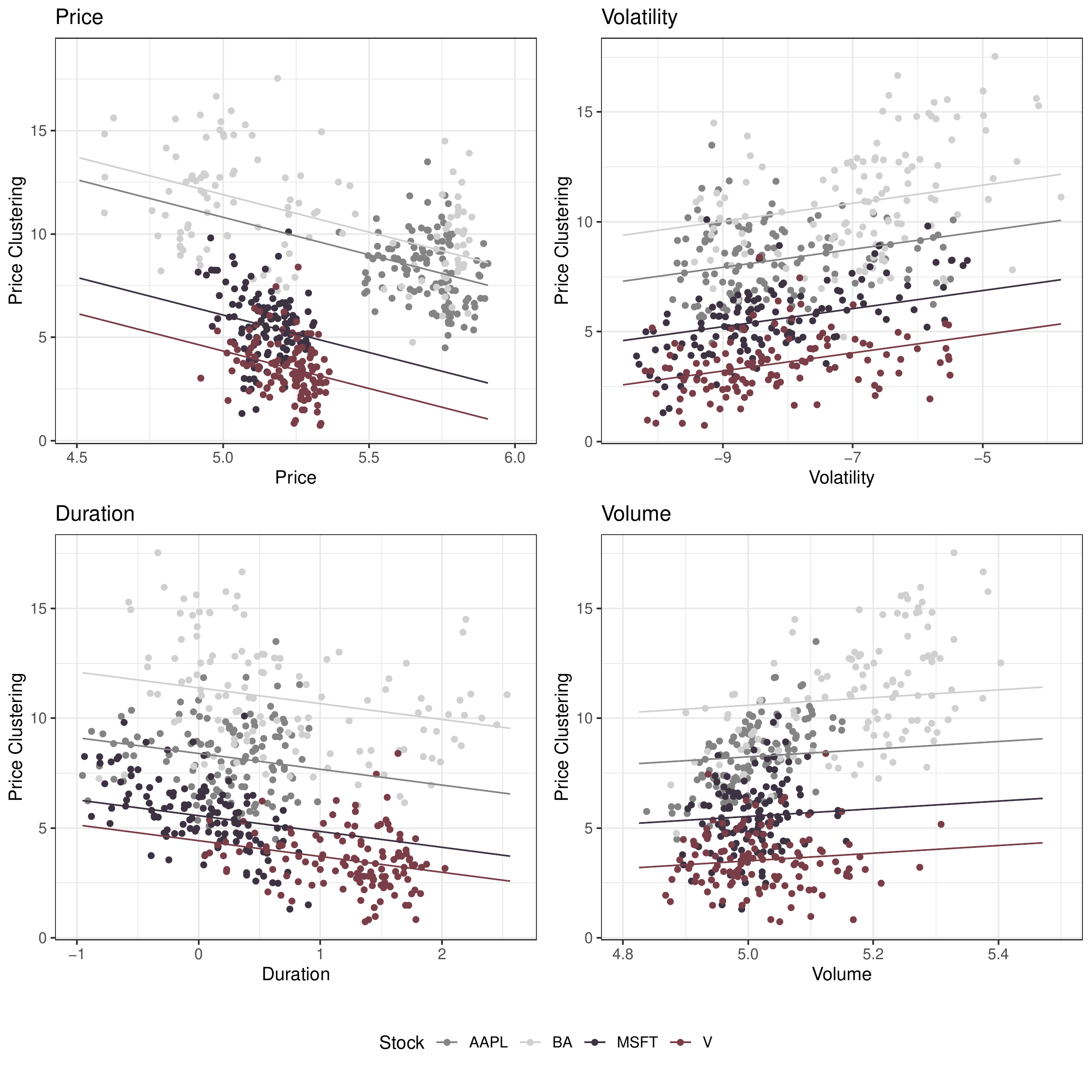} 
\caption{Daily data of four selected stocks and fitted lines from univariate panel regressions of price clustering on logarithmic price (top left), logarithmic volatility (top right), logarithmic duration (bottom left) and logarithmic volume (bottom right). All models are estimated with stock fixed effects.}
\label{fig:daily}
\end{center}
\end{figure}

Table \ref{tab:dailyCoef} reports estimated coefficients of three variants of the fixed effects model. The first variant models price clustering on price, volatility and duration, i.e. model in \eqref{eq:plm} where volume is skipped. The second model considers only the price, duration and volume as the explanatory variables, and the third one is the full model in \eqref{eq:plm}. We test the significance of the estimated coefficients using robust standard errors for which observations are clustered in both dimensions to account for serial as well as cross-sectional correlation. The results show that only volatility and volume are significant drivers of the price clustering in the full model (Model III). However, once the volatility is dropped from the model, the average daily duration becomes highly significant (Model II). A similar result applies for the price (Model I) which becomes highly significant once the volume is dropped from the full model. For illustration, Figure \ref{fig:daily} shows fitted lines from univariate regressions with stock specific effects\footnote{Time effects are dropped for better visibility which does not alter the main result.} for two stocks traded on NASDAQ -- Apple Inc. (AAPL) and Microsoft (MSFT) -- and two stocks traded on NYSE -- Boeing (BA) and Visa Inc. (V). 

\begin{table}[!htbp] \centering 
  \caption{Estimated coefficients and robust standard errors of models with fixed effects for stocks and days. Label I refers to the model in \eqref{eq:plm} for which the volume parameter $\beta_4 = 0$; II to the model in \eqref{eq:plm} for which the volatility parameter $\beta_2 = 0$; III to the full model in \eqref{eq:plm}.} 
  \label{tab:dailyCoef} 
\begin{tabular}{@{\extracolsep{5pt}}lD{.}{.}{-4} D{.}{.}{-4} D{.}{.}{-4} } 
\\ \toprule
Variable & \multicolumn{1}{c}{I} & \multicolumn{1}{c}{II} & \multicolumn{1}{c}{III}\\ 
\midrule \\[-1.8ex]
 price & -1.7014^{***} & -0.8067 & -0.1245 \\ 
  & (0.5469) & (0.6615) & (0.5323) \\ 
  & & & \\ 
 volatility & 0.5900^{***} &  & 0.7008^{***} \\ 
  & (0.1901) &  & (0.1622) \\ 
  & & & \\ 
 duration & -0.0973 & -0.4640^{***} & -0.0069 \\ 
  & (0.1672) & (0.1347) & (0.1739) \\ 
  & & & \\ 
 volume &  & 3.7544^{***} & 3.9557^{***} \\ 
  &  & (0.4869) & (0.5025) \\ 
  & & & \\ [-1.8ex] 
\bottomrule \\[-1.8ex] 
  & \multicolumn{3}{r}{$^{*}$p$<$0.05; $^{**}$p$<$0.01; $^{***}$p$<$0.001} \\ 
\end{tabular} 
\end{table}  

\subsection{High-Frequency Analysis}
\label{sec:empHigh}

\begin{figure}
\begin{center}
\includegraphics[width=15cm]{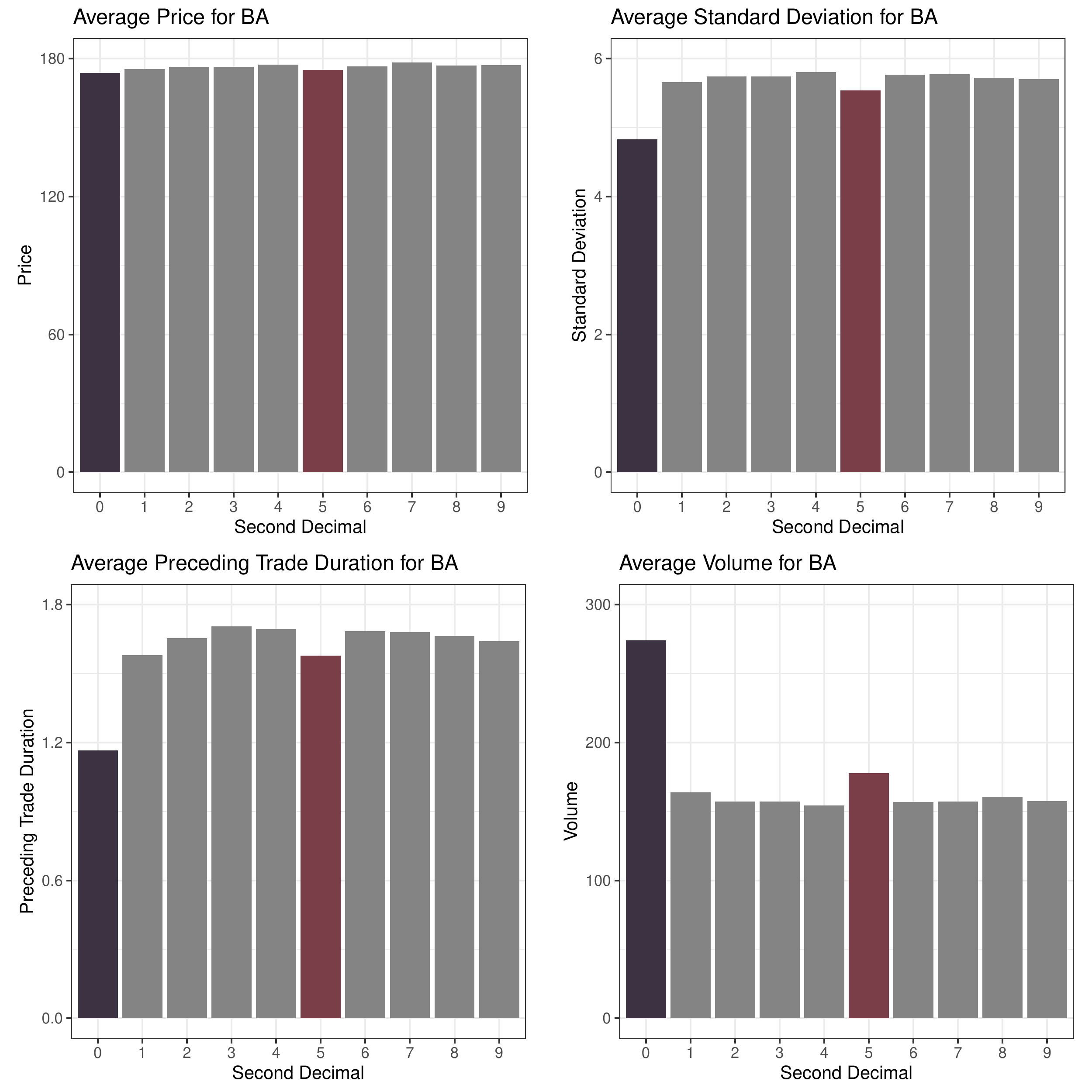} 
\caption{The price (top left), the average standard deviation (top right), the average preceding trade duration (bottom left), and the volume (bottom right) broken down by the second decimal digit of the BA stock prices.}
\label{fig:average}
\end{center}
\end{figure}

Let us analyze the price clustering phenomenon at the highest possible frequency. First, we take a brief look at the relation between the individual explanatory variables and price clustering. We focus on the BA stock as its price clustering is the most pronounced. Figure \ref{fig:average} shows the average expected price, the average instantaneous variance obtained from the dynamic model, the average duration preceding the trade, and the average volume broken down by the second decimal of the price for the BA stock. We can clearly see that for prices ending with 0 and 5, the average variance and the average trade duration is much lower than for the other digits while the average volume is much higher. Note that succeeding durations show very similar behavior to preceding durations suggesting that price clustering tends to occur when trading is more intense.

Next, we estimate three versions of the proposed price clustering model for each of the 30 stocks. In the first version, we assume that there is no price clustering and set $f = g_1 = g_2 = g_3 = g_4 = h_{5} = h_{10} = 0$. In the second version, we set only $f = g_1 = g_2 = g_3 = g_4 = 0$ and assume that there is price clustering present but is constant over time and does not depend on any variables. The third version is the dynamic model presented in Section \ref{sec:modelDynamics} without any restrictions. We report the average log-likelihood and the Akaike information criterion (AIC) of the models in Table \ref{tab:highAic}. We can see that adding price clustering to the model and subsequently adding dynamics to price clustering is very much worth of the extra few parameters as AIC is distinctly the lowest for the dynamic model for all stocks.

From now on we focus on the model with dynamic price clustering. Table \ref{tab:highCoef} reports the estimated coefficients. For all stocks, the coefficients in the volatility process $c$, $b$, $a$, and $d$ have the same signs and fairly similar values demonstrating the robustness of the model. Parameter $d$ is negative, which means that with longer durations, dispersion parameter $\alpha_t$ is lower and the instantaneous variance $\mu_t \exp (-\alpha_t)$ is higher. We attribute this behavior to the presence of a large amount of extremely short durations associated with small price changes. Note that for example, the BA stock has 50 percent of durations shorter than 0.1 seconds and 19 percent shorter than 0.0001 seconds. \cite{Engle2000} observed the opposite relation between trade durations and volatility but based his results on data with a much lower frequency and without durations shorter than 1 second. This might indicate a change in the data structure over the years and a complex non-linear relation between trade durations and volatility. This topic is, however, beyond the scope of this paper.

Regarding the dynamics of price clustering, the autoregressive parameter $f$ is stable across all stocks. The parameter for the expected price $g_1$ significantly varies for different stocks suggesting its low informative value. This is in line with the daily analysis in which prices were found insignificant. The parameter for the instantaneous variance $g_2$ is positive for all stocks. The portion of one cent traders is therefore higher with higher variance and price clustering tends to occur when prices are less volatile. This is the most interesting result as it deviates from the behavior observed in the daily analysis. The parameter for the preceding trade duration $g_3$ significantly varies for different stocks, similarly to $g_1$. When the preceding trade duration is the only explanatory variable included in the model, however, $g_3$ is positive for all stocks. Recall that durations have a positive effect on instantaneous variance as $d$ is positive for all stocks. This implies that durations have an effect on instantaneous variance which in turn has an effect on price clustering. However, when controlling for instantaneous variance, durations do not bring additional information to explain price clustering. These observations are in line with the daily analysis. Finally, the parameter for the volume $g_4$ is negative for all stocks. As in the daily analysis, higher volume is clearly associated with higher price clustering.

\begin{figure}
\begin{center}
\includegraphics[width=15cm]{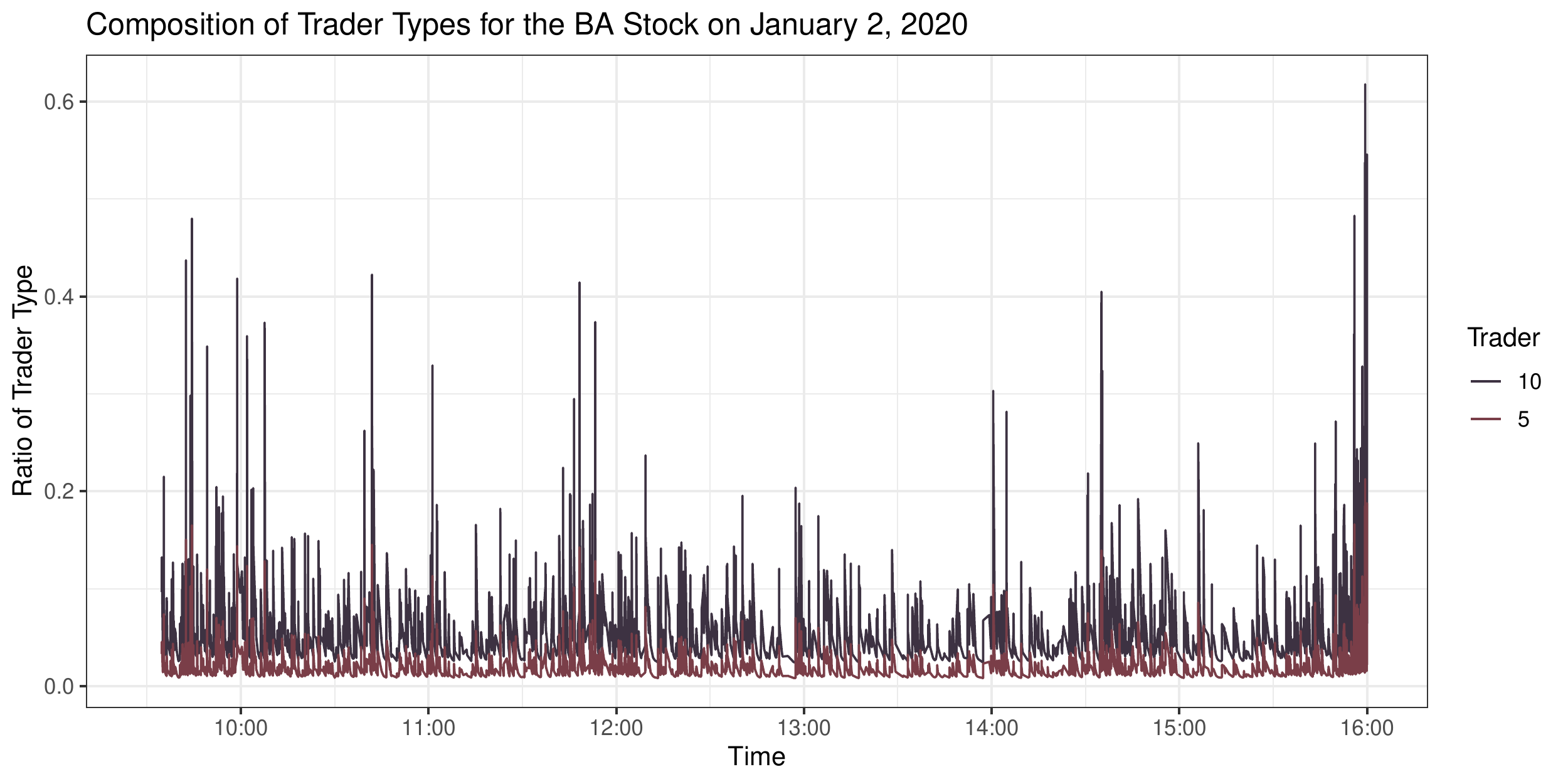} 
\caption{The time-varying portions of trader types obtained from the proposed price clustering model for the BA stock on January 2, 2020.}
\label{fig:time}
\end{center}
\end{figure}

We omit parameters $h_5$ and $h_{10}$ controlling strength of price clustering from Table \ref{tab:highCoef} as they are not very informative for readers. It is far better to look at the average values of trader portions $\bar{\varphi}_{1}$, $\bar{\varphi}_{5}$ and $\bar{\varphi}_{10}$ reported in Table \ref{tab:highUnc}. The average portion of ten-cent traders ranges from 0.48 percent for the TRV stock to 10.54 percent for the BA stock. The average portion of five-cent traders ranges from 0.34 percent for the TRV stock to 3.63 percent for the BA stock. An example of the progression of trader type ratios is shown in Figure \ref{fig:time} for the BA stock on the first trading day of 2020.

\begin{table}
\centering
\caption{The average log-likelihood and AIC of the model without price clustering, the model with static price clustering, and the model with dynamic price clustering.}
\label{tab:highAic}
\begin{tabular}{lrrrrrr}
\toprule
& \multicolumn{2}{c}{No PC} & \multicolumn{2}{c}{Static PC} & \multicolumn{2}{c}{Dynamic PC} \\ 
Stock & Lik. & AIC & Lik. & AIC & Lik. & AIC \\ 
\midrule
AAPL & -2.3548 & \num{12080434} & -2.3292 & \num{11949022} & -2.2953 & \num{11775068} \\ 
AXP  & -2.5117 &  \num{2284414} & -2.5054 &  \num{2278659} & -2.5008 &  \num{2274477} \\ 
BA   & -2.8677 & \num{10745431} & -2.8146 & \num{10546574} & -2.7980 & \num{10484266} \\ 
CAT  & -2.6144 &  \num{2092986} & -2.6090 &  \num{2088625} & -2.6066 &  \num{2086695} \\ 
CSCO & -0.4650 &  \num{1525535} & -0.4636 &  \num{1520852} & -0.4606 &  \num{1510984} \\ 
CVX  & -2.1513 &  \num{4094545} & -2.1476 &  \num{4087384} & -2.1444 &  \num{4081369} \\ 
DIS  & -2.1382 &  \num{5620020} & -2.1295 &  \num{5597162} & -2.1231 &  \num{5580359} \\ 
DOW  & -1.6701 &  \num{1981601} & -1.6685 &  \num{1979699} & -1.6668 &  \num{1977626} \\ 
GS   & -3.1560 &  \num{2292436} & -3.1500 &  \num{2288064} & -3.1460 &  \num{2285185} \\ 
HD   & -3.1018 &  \num{3043787} & -3.0969 &  \num{3039026} & -3.0935 &  \num{3035655} \\ 
IBM  & -2.4432 &  \num{2689846} & -2.4393 &  \num{2685501} & -2.4360 &  \num{2681876} \\ 
INTC & -0.8455 &  \num{3165039} & -0.8442 &  \num{3160026} & -0.8407 &  \num{3147091} \\ 
JNJ  & -2.3836 &  \num{3976323} & -2.3809 &  \num{3971919} & -2.3794 &  \num{3969402} \\ 
JPM  & -2.0310 &  \num{5831271} & -2.0264 &  \num{5817872} & -2.0220 &  \num{5805430} \\ 
KO   & -1.2238 &  \num{2758822} & -1.2230 &  \num{2757081} & -1.2212 &  \num{2753113} \\ 
MCD  & -2.9611 &  \num{2787186} & -2.9563 &  \num{2782675} & -2.9526 &  \num{2779159} \\ 
MMM  & -2.7229 &  \num{2356746} & -2.7180 &  \num{2352455} & -2.7149 &  \num{2349844} \\ 
MRK  & -1.7024 &  \num{3763397} & -1.7011 &  \num{3760622} & -1.6998 &  \num{3757820} \\ 
MSFT & -1.7431 & \num{10325970} & -1.7315 & \num{10257392} & -1.7096 & \num{10127509} \\ 
NKE  & -2.1331 &  \num{2878657} & -2.1310 &  \num{2875789} & -2.1292 &  \num{2873371} \\ 
PFE  & -0.8423 &  \num{2403152} & -0.8415 &  \num{2400862} & -0.8398 &  \num{2396048} \\ 
PG   & -2.3049 &  \num{3882847} & -2.3024 &  \num{3878712} & -2.3007 &  \num{3875756} \\ 
RTX  & -1.7279 &  \num{1602691} & -1.7238 &  \num{1598954} & -1.7207 &  \num{1596096} \\ 
TRV  & -2.7871 &  \num{1135714} & -2.7861 &  \num{1135292} & -2.7848 &  \num{1134797} \\ 
UNH  & -3.3882 &  \num{3716005} & -3.3834 &  \num{3710792} & -3.3816 &  \num{3708786} \\ 
V    & -2.6542 &  \num{5058066} & -2.6485 &  \num{5047243} & -2.6434 &  \num{5037606} \\ 
VZ   & -1.3165 &  \num{3025851} & -1.3156 &  \num{3023684} & -1.3141 &  \num{3020201} \\ 
WBA  & -1.1807 &  \num{1507681} & -1.1787 &  \num{1505148} & -1.1748 &  \num{1500227} \\ 
WMT  & -2.1291 &  \num{3718970} & -2.1262 &  \num{3714030} & -2.1233 &  \num{3708967} \\ 
XOM  & -1.1969 &  \num{4893716} & -1.1951 &  \num{4886388} & -1.1895 &  \num{4863486} \\ 
\bottomrule
\end{tabular}
\end{table}

\begin{table}
\centering
\caption{Estimated coefficients of the proposed dynamic price clustering model.}
\label{tab:highCoef}
\begin{tabular}{lrrrrrrrrr}
\toprule
Stock & $c$ & $b$ & $a$ & $d$ & $f$ & $g_1$ & $g_2$ & $g_3$ & $g_4$ \\ 
\midrule
AAPL & 6.09 & 0.08 & 0.29 & -0.40 & 0.66 & -0.33 & 0.61 & -0.11 & -0.69 \\ 
AXP & 5.32 & 0.10 & 0.35 & -0.28 & 0.34 & 1.05 & 0.05 & 0.06 & -0.74 \\ 
BA & 5.00 & 0.09 & 0.30 & -0.29 & 0.39 & -0.14 & 0.18 & 0.03 & -0.71 \\ 
CAT & 5.60 & 0.04 & 0.26 & -0.27 & 0.25 & -0.28 & 0.29 & -0.00 & -0.71 \\ 
CSCO & 6.11 & 0.17 & 0.08 & -0.35 & 0.81 & 0.28 & 0.40 & -0.08 & -0.37 \\ 
CVX & 5.79 & 0.11 & 0.35 & -0.26 & 0.36 & 0.70 & 0.28 & 0.02 & -0.64 \\ 
DIS & 6.05 & 0.12 & 0.33 & -0.25 & 0.39 & 0.45 & 0.21 & 0.05 & -0.64 \\ 
DOW & 5.71 & 0.15 & 0.33 & -0.22 & 0.34 & 0.55 & 0.18 & 0.04 & -0.70 \\ 
GS & 4.89 & 0.05 & 0.27 & -0.31 & 0.29 & 1.06 & 0.12 & 0.02 & -0.90 \\ 
HD & 5.01 & 0.08 & 0.29 & -0.28 & 0.29 & 0.64 & 0.10 & 0.05 & -0.83 \\ 
IBM & 5.79 & 0.08 & 0.32 & -0.28 & 0.32 & 0.33 & 0.10 & 0.06 & -0.80 \\ 
INTC & 6.29 & 0.14 & 0.13 & -0.35 & 0.74 & 0.38 & 0.34 & -0.06 & -0.63 \\ 
JNJ & 5.76 & 0.13 & 0.40 & -0.25 & 0.23 & 1.12 & 0.08 & 0.05 & -0.74 \\ 
JPM & 5.76 & 0.17 & 0.44 & -0.26 & 0.33 & 0.48 & 0.67 & -0.07 & -0.56 \\ 
KO & 5.97 & 0.25 & 0.33 & -0.18 & 0.55 & -0.05 & 0.62 & 0.01 & -0.43 \\ 
MCD & 5.16 & 0.08 & 0.34 & -0.27 & 0.31 & 1.69 & 0.02 & 0.06 & -0.88 \\ 
MMM & 5.55 & 0.05 & 0.29 & -0.27 & 0.25 & 0.93 & 0.06 & 0.06 & -0.87 \\ 
MRK & 5.89 & 0.20 & 0.40 & -0.23 & 0.45 & 0.20 & 0.71 & -0.09 & -0.53 \\ 
MSFT & 6.37 & 0.11 & 0.27 & -0.38 & 0.72 & 0.17 & 0.39 & -0.05 & -0.72 \\ 
NKE & 6.01 & 0.10 & 0.35 & -0.25 & 0.29 & 0.45 & 0.05 & 0.08 & -0.70 \\ 
PFE & 5.19 & 0.38 & 0.28 & -0.15 & 0.70 & 0.36 & 0.34 & 0.03 & -0.33 \\ 
PG & 5.84 & 0.11 & 0.40 & -0.23 & 0.41 & -0.13 & 1.35 & -0.25 & -0.53 \\ 
RTX & 6.12 & 0.15 & 0.34 & -0.24 & 0.32 & -0.50 & 0.02 & 0.10 & -0.65 \\ 
TRV & 5.07 & 0.05 & 0.30 & -0.30 & 0.96 & -0.66 & 0.57 & -0.14 & -0.02 \\ 
UNH & 4.58 & 0.07 & 0.28 & -0.27 & 0.39 & 0.08 & 0.11 & 0.03 & -0.73 \\ 
V & 5.91 & 0.06 & 0.32 & -0.27 & 0.32 & 0.65 & 0.09 & 0.06 & -0.91 \\ 
VZ & 5.79 & 0.27 & 0.38 & -0.19 & 0.44 & 2.53 & 0.71 & -0.05 & -0.53 \\ 
WBA & 5.82 & 0.12 & 0.18 & -0.34 & 0.68 & 0.38 & 0.26 & -0.00 & -0.62 \\ 
WMT & 6.19 & 0.12 & 0.38 & -0.24 & 0.34 & -2.33 & 0.15 & 0.08 & -0.71 \\ 
XOM & 6.11 & 0.24 & 0.31 & -0.18 & 0.57 & -0.51 & 0.98 & -0.06 & -0.52 \\ 
\bottomrule
\end{tabular}
\end{table}

\begin{table}
\centering
\caption{The average values of the time-varying parameters of the proposed dynamic price clustering model. Values of $\bar{\mu}$ are in dollars and values of $\bar{\varphi}_1$, $\bar{\varphi}_5$, and $\bar{\varphi}_{10}$ are in percent.}
\label{tab:highUnc}
\begin{tabular}{lrrrrr}
\toprule
Stock & $\bar{\mu}$ & $\bar{\alpha}$ & $\bar{\varphi}_1$ & $\bar{\varphi}_5$ & $\bar{\varphi}_{10}$ \\ 
\midrule
AAPL & 291.80 & 8.43 & 92.37 & 0.72 & 6.91 \\ 
  AXP & 98.43 & 7.01 & 95.05 & 1.81 & 3.15 \\ 
  BA & 175.80 & 6.85 & 85.84 & 3.63 & 10.54 \\ 
  CAT & 117.91 & 6.99 & 95.40 & 1.85 & 2.75 \\ 
  CSCO & 42.22 & 10.26 & 98.07 & 0.54 & 1.38 \\ 
  CVX & 88.26 & 7.62 & 96.26 & 1.32 & 2.42 \\ 
  DIS & 112.50 & 7.90 & 94.18 & 2.25 & 3.56 \\ 
  DOW & 36.46 & 7.69 & 97.36 & 1.36 & 1.29 \\ 
  GS & 193.14 & 6.40 & 95.49 & 1.06 & 3.45 \\ 
  HD & 215.44 & 6.63 & 95.72 & 1.44 & 2.84 \\ 
  IBM & 124.05 & 7.38 & 96.01 & 1.64 & 2.34 \\ 
  INTC & 57.72 & 9.82 & 97.83 & 0.65 & 1.52 \\ 
  JNJ & 140.37 & 7.63 & 96.78 & 1.36 & 1.85 \\ 
  JPM & 103.65 & 8.03 & 95.47 & 2.54 & 1.98 \\ 
  KO & 48.33 & 8.88 & 98.29 & 0.67 & 1.04 \\ 
  MCD & 183.86 & 6.74 & 96.12 & 0.72 & 3.16 \\ 
  MMM & 149.92 & 7.02 & 95.87 & 1.17 & 2.96 \\ 
  MRK & 79.28 & 8.43 & 97.69 & 1.23 & 1.08 \\ 
  MSFT & 169.29 & 9.11 & 94.57 & 0.84 & 4.59 \\ 
  NKE & 88.58 & 7.67 & 96.89 & 1.60 & 1.51 \\ 
  PFE & 34.46 & 9.30 & 98.07 & 1.01 & 0.91 \\ 
  PG & 116.41 & 7.61 & 96.82 & 1.87 & 1.31 \\ 
  RTX & 62.28 & 8.13 & 95.83 & 2.06 & 2.11 \\ 
  TRV & 111.57 & 6.59 & 99.18 & 0.34 & 0.48 \\ 
  UNH & 271.73 & 6.29 & 96.01 & 1.08 & 2.91 \\ 
  V & 180.13 & 7.34 & 95.59 & 1.03 & 3.37 \\ 
  VZ & 55.60 & 8.85 & 97.90 & 1.13 & 0.97 \\ 
  WBA & 45.73 & 8.91 & 97.49 & 0.72 & 1.79 \\ 
  WMT & 118.65 & 7.98 & 96.57 & 1.56 & 1.87 \\ 
  XOM & 45.82 & 8.86 & 97.02 & 1.54 & 1.44 \\ 
\bottomrule
\end{tabular}
\end{table}

\subsection{Implications}
\label{sec:empImp}

Several hypotheses have been established to explain the price clustering phenomenon. The \emph{attraction hypothesis} of \cite{Goodhart1991} essentially states that there exists a particular preference (basic attraction) for certain numbers, especially for the rounded ones. The \emph{negotiation hypothesis} of \cite{Harris1991} assumes that traders use discrete price sets to lower the costs of negotiating. Once the set of prices is reduced, the traders reach agreements more easily since the amount of information that must be exchanged between negotiating traders decreases. \cite{Christie1994a} argued in the \emph{collusion hypothesis} that the lack of odd-eighth quotes on NASDAQ cannot be explained by the negotiation hypothesis, trading activity, or other variables thought to impact spreads, which suggests that NASDAQ dealers might implicitly collude to maintain wide spreads. However, assessing these hypotheses is out of the scope of our paper. In this section, we focus only on the most studied hypothesis in the literature -- the price resolution hypothesis.

The \emph{price resolution hypothesis} of \cite{Ball1985} considers the source of price clustering to be the uncertainty. It states that when the amount of information in the market is low and the volatility becomes higher, the market participants incline to round their prices, and consequently, the price clustering increases. This hypothesis was confirmed by many studies. The studies found that price clustering increases with volatility using different data and measures. For example, \cite{Ahn2005} computed the volatility as the inverse of the daily return standard deviation, while \cite{Ikenberry2008} used the standard deviation of returns over the sample period, \cite{Box2016} used the standard deviation of 15-minute continuously compounded midpoint returns over the trading day, \cite{Schwartz2004} used the difference in the high and low prices for the day, and \cite{Lien2019} utilized the transitory volatility defined as the coefficient of variation of intraday trade prices. \cite{Davis2014} found that price clustering is positively related to volatility, however, only when a non–high-frequency trading firm provides liquidity. On the contrary to the vast majority of the literature, \cite{Blau2019} reported based on panel regressions that the volatility is negatively related to price clustering, where the volatility is measured as the standard deviation of residual returns obtained from estimating a Fama and French 3-factor model.

Our results from the daily analysis show that the realized volatility is highly significant and positively related to the price clustering. This finding is in line with the price resolution hypothesis. Interestingly, instantaneous volatility obtained from the proposed dynamic price model has a negative effect on price clustering. The results do not contradict since they explore price clustering from different perspectives. The result based on daily data holds for low-frequency traders whose price resolution is influenced by the uncertainty in a negative way, i.e. the higher daily volatility, the higher price clustering. On the other hand, the presence of high-frequency traders is typically associated with increased volatility (see, e.g., \citealp{Rosu2019, Shkilko2020, Boehmer2020}). Moreover, high-frequency traders generally do not incline to price rounding (see \citealp{Davis2014}). Consequently, the higher the instantaneous volatility is, the higher portion of high-frequency traders is, which lowers the price clustering.

\section{Conclusion}
\label{sec:conclusion}

We have proposed a dynamic price model to capture agents trading in different multiples of the tick size. In the literature, this empirical phenomenon known as price clustering was mostly approached only by basic descriptive statistics rather than a proper price model. By analyzing 30 DJIA stocks from both daily and high-frequency perspectives, we have revealed dissension between the two time scales. While daily realized volatility has a positive effect on price clustering, instantaneous volatility obtained by the proposed model has a negative effect. We argue that volatility on lower frequency affects low-frequency traders through the resolution hypothesis while volatility on higher frequency affects only high-frequency traders who do not tend to price clustering.

We believe the model to be sufficient for its purpose -- capturing price clustering and allowing to explain it. For the model to be able to compete with other high-frequency price models, however, it would have to be improved. The main limitation lies in the underlying distribution. We have yet to study how well the double Poisson distribution, which we have used, captures the observed prices. However, due to our specific problem, we require the distribution to be defined on positive integers and allow for underdispersion. The range of possible alternatives is therefore severely limited as it is not a typical situation in count data analysis. Furthermore, the specification of the dynamics could be enhanced. We could include a separate model for durations and we could add a seasonality component to the volatility. One possible direction for the future research is therefore to assess the suitability of the double Poisson distribution for prices, extend the specification of the proposed dynamic model and compare it with various models for price differences.

Concerning the empirical study, our focus has been on the price variance, whether it is daily realized volatility or instantaneous variance. Nevertheless, we have also included the expected price, the preceding trade duration, and the volume as explanatory variables. These are the most common variables in the price clustering literature. However, other factors such as the spread and the investor sentiment could also be considered. In the context of the proposed high-frequency price model, any variable could be straightforwardly included in the price clustering dynamics. Analyzing the effects of these factors is the second possible direction of the future research.

\section*{Acknowledgements}
\label{sec:acknow}

Computational resources were supplied by the project "e-Infrastruktura CZ" (e-INFRA LM2018140) provided within the program Projects of Large Research, Development and Innovations Infrastructures.

\section*{Funding}
\label{sec:fund}

This research was supported by the Czech Science Foundation under project 19-02773S, the Internal Grant Agency of the Prague University of Economics and Business under project F4/53/2019, and the Institutional Support Funds for the long-term conceptual development of the Faculty of Informatics, Prague University of Economics and Business.

\appendix

\section{Derivation of Distribution for Specific Trader Types}
\label{app:mixture}

Let there be $m$ types of traders that can trade only in $k_1, \ldots, k_m$ multiples of the tick size respectively. For trader type $k \in \{ k_1, \ldots, k_m \}$, we derive the distribution of prices $\mathrm{P} \left[ Y^{[k]} = y \middle| \mu, \alpha \right]$. We require the distribution to be based on the double Poisson distribution, to have the support consisting of multiples of $k$, to have the expected value $\mathrm{E} \left[ Y^{[k]} \right] \simeq \mu$ and to have the variance $\mathrm{var} \left[ Y^{[k]} \right] \simeq \mu e^{-\alpha}$. We can modify any integer distribution $\mathrm{P} \left[ Z^{[k]} = y \middle| \mu, \alpha \right]$ to have support consisting only of multiples of $k$ as
\begin{equation}
\mathrm{P} \left[ Y^{[k]} = y \middle| \mu, \alpha \right] = \mathbb{I} \left\{ k \mid y \right\} \mathrm{P} \left[ Z^{[k]} = \frac{y}{k} \middle| \mu, \alpha \right],
\end{equation}
where $\mathbb{I} \left\{ k \mid y \right\}$ is equal to 1 if $y$ is divisible by $k$ and 0 otherwise. We assume that $Z^{[k]}$ follows the double Poisson distribution with parameters $ \mu^{[k]}$ and $\alpha^{[k]}$, i.e.\ $Z^{[k]} \sim \textrm{DP} \left( \mu^{[k]}, \alpha^{[k]} \right)$. The expected value of $Y^{[k]}$ is
\begin{equation}
\begin{aligned}
\mathrm{E} \left[ Y^{[k]} \right] &= \sum_{y=0}^{\infty} y \mathrm{P} \left[ Y^{[k]} = y \middle| \mu, \alpha \right] \\
&= \sum_{y=0}^{\infty} y \mathbb{I} \left\{ k \mid y \right\} \mathrm{P} \left[ Z^{[k]} = \frac{y}{k} \middle| \mu, \alpha \right] \\
&= \sum_{y=0}^{\infty} k y \mathrm{P} \left[ Z^{[k]} = y \middle| \mu, \alpha \right] \\
&= k \mathrm{E} \left[ Z^{[k]} \right] \\
&\simeq k \mu^{[k]}.
\end{aligned}
\end{equation}
The variance of $Y^{[k]}$ is
\begin{equation}
\begin{aligned}
\mathrm{var} \left[ Y^{[k]} \right] &= \sum_{y=0}^{\infty} \left( y - \mathrm{E} \left[ Y^{[k]} \right] \right)^2 \mathrm{P} \left[ Y^{[k]} = y \middle| \mu, \alpha \right] \\
&= \sum_{y=0}^{\infty} \left( y - \mathrm{E} \left[ Y^{[k]} \right] \right)^2 \mathbb{I} \left\{ k \mid y \right\} \mathrm{P} \left[ Z^{[k]} = \frac{y}{k} \middle| \mu, \alpha \right] \\
&= \sum_{y=0}^{\infty} \left( k y - k \mathrm{E} \left[ Z^{[k]} \right] \right)^2 \mathrm{P} \left[ Z^{[k]} = y \middle| \mu, \alpha \right] \\
&= k^2 \mathrm{var} \left[ Z^{[k]} \right] \\
&\simeq k^2 \mu^{[k]} e^{-\alpha^{[k]}}.
\end{aligned}
\end{equation}
Our last requirements $\mathrm{E} \left[ Y^{[k]} \right] \simeq \mu$ with $\mathrm{var} \left[ Y^{[k]} \right] \simeq \mu e^{-\alpha}$ lead to the system of equations
\begin{equation}
\begin{aligned}
\mu &= k \mu^{[k]} \\
\mu e^{-\alpha} &= k^2 \mu^{[k]} e^{-\alpha^{[k]}}
\end{aligned}
\end{equation}
with the solution
\begin{equation}
\mu^{[k]} = \frac{\mu}{k}, \qquad \alpha^{[k]} = \alpha + \ln \left( k \right). \\
\end{equation}
Everything together gives us the distribution
\begin{equation}
\mathrm{P} \left[ Y^{[k]} = y \middle| \mu, \alpha \right] = \mathbb{I} \left\{ k \mid y \right\} \mathrm{P} \left[ Z^{[k]} = \frac{y}{k} \middle| \mu, \alpha \right], \qquad Z^{[k]} \sim \mathrm{DP} \left( \frac{\mu}{k}, \alpha + \ln(k) \right).
\end{equation}
Note that the mixture distribution of all prices
\begin{equation}
\mathrm{P} \left[ Y = y \middle| \mu, \alpha, \varphi_{k_1}, \ldots, \varphi_{k_m} \right] = \sum_{k \in \{ k_1, \ldots, k_m \}} \varphi_{k} \mathrm{P} \left[ Y^{[k]} = y \middle| \mu, \alpha \right]
\end{equation}
has approximately the same expected value and variance as the distribution of $Y^{[k]}$. This is based on the identity
\begin{equation}
\begin{aligned}
\mathrm{E} \left[ g(Y) \right] &= \sum_{y=0}^{\infty} g(y) \mathrm{P} \left[ Y = y \middle| \mu, \alpha, \varphi_{k_1}, \ldots, \varphi_{k_m} \right] \\
&= \sum_{y=0}^{\infty} g(y) \sum_{k \in \{ k_1, \ldots, k_m \}} \varphi_{k} \mathrm{P} \left[ Y^{[k]} = y \middle| \mu, \alpha \right] \\
&= \sum_{k \in \{ k_1, \ldots, k_m \}} \varphi_{k} \sum_{y=0}^{\infty} g(y) \mathrm{P} \left[ Y^{[k]} = y \middle| \mu, \alpha \right] \\
&= \sum_{k \in \{ k_1, \ldots, k_m \}} \varphi_{k} \mathrm{E} \left[ g \left( Y^{[k]} \right) \right] \\
&= \mathrm{E} \left[ g \left( Y^{[k]} \right) \right],
\end{aligned}
\end{equation}
where $g(\cdot)$ is any function satisfying that $\mathrm{E} \left[ g \left( Y^{[k]} \right) \right]$ are the same for all $k$.

\newpage
\section{Descriptive Statistics of Cleaned Data}
\label{app:descriptive}

\begin{table}[ht!]
\centering
\caption{The table reports a number of observations (\#Trades), sample mean (Mean P) and standard deviation (SD P) of prices, sample mean (Mean D) and standard deviation (SD D) of durations, and price clustering (PC) calculated as the excess relative frequency of multiples of five cents and ten cents in prices.}
\label{tab:descriptive}
\begin{tabular}{lrrrrrr}
\toprule
Stock & \#Trades       & Mean P       & SD P        & Mean D        & SD D         & PC [\%]  \\
\midrule
AAPL  & \num{2671590} & \num{291.71} & \num{34.67} & \num{1.10}  & \num{2.71}  & \num{8.28}  \\
AXP   & \num{467623}  & \num{98.59}  & \num{17.84} & \num{6.24}  & \num{13.30} & \num{4.02}  \\
BA    & \num{1886402} & \num{176.06} & \num{60.40} & \num{1.55}  & \num{5.01}  & \num{11.52} \\
CAT   & \num{413148}  & \num{118.07} & \num{14.32} & \num{7.08}  & \num{15.25} & \num{3.79}  \\
CSCO  & \num{1712341} & \num{42.26}  & \num{4.14}  & \num{1.71}  & \num{5.54}  & \num{2.08}  \\
CVX   & \num{964508}  & \num{88.34}  & \num{15.87} & \num{3.03}  & \num{6.42}  & \num{3.03}  \\
DIS   & \num{1327041} & \num{112.56} & \num{16.38} & \num{2.20}  & \num{4.67}  & \num{4.80}  \\
DOW   & \num{606116}  & \num{36.53}  & \num{8.33}  & \num{4.82}  & \num{10.44} & \num{2.24}  \\
GS    & \num{376058}  & \num{193.37} & \num{31.28} & \num{7.78}  & \num{16.89} & \num{3.53}  \\
HD    & \num{503522}  & \num{215.65} & \num{28.69} & \num{5.80}  & \num{12.12} & \num{3.57}  \\
IBM   & \num{563345}  & \num{124.12} & \num{15.30}  & \num{5.19}  & \num{10.60} & \num{3.34}  \\
INTC  & \num{2065813} & \num{57.84}  & \num{5.53}  & \num{1.42}  & \num{4.22}  & \num{2.13}  \\
JNJ   & \num{846988}  & \num{140.42} & \num{9.87}  & \num{3.45}  & \num{7.14}  & \num{2.68}  \\
JPM   & \num{1448425} & \num{103.69} & \num{17.38} & \num{2.02}  & \num{4.30}  & \num{3.71}  \\
KO    & \num{1140050} & \num{48.35}  & \num{5.78}  & \num{2.56}  & \num{6.63}  & \num{1.45}  \\
MCD   & \num{483508}  & \num{184.02} & \num{20.76} & \num{6.05}  & \num{12.52} & \num{3.16}  \\
MMM   & \num{445633}  & \num{150.04} & \num{14.01} & \num{6.56}  & \num{13.99} & \num{3.51}  \\
MRK   & \num{1118215} & \num{79.30}  & \num{5.40}  & \num{2.61}  & \num{5.55}  & \num{1.94}  \\
MSFT  & \num{3099279} & \num{169.25} & \num{16.27} & \num{0.94}  & \num{2.40}  & \num{5.96}  \\
NKE   & \num{687619}  & \num{88.66}  & \num{11.31} & \num{4.25}  & \num{8.78}  & \num{2.50}  \\
PFE   & \num{1439433} & \num{34.48}  & \num{3.06}  & \num{2.03}  & \num{5.73}  & \num{1.56}  \\
PG    & \num{855186}  & \num{116.44} & \num{6.85}  & \num{3.41}  & \num{7.14}  & \num{2.79}  \\
RTX   & \num{470061}  & \num{62.28}  & \num{4.87}  & \num{3.04}  & \num{5.49}  & \num{3.37}  \\
TRV   & \num{216618}  & \num{111.75} & \num{17.08} & \num{13.49} & \num{29.25} & \num{1.84}  \\
UNH   & \num{561256}  & \num{271.93} & \num{26.72} & \num{5.21}  & \num{12.16} & \num{3.39}  \\
V     & \num{965718}  & \num{180.20} & \num{18.36} & \num{3.03}  & \num{6.07}  & \num{3.68}  \\
VZ    & \num{1162040} & \num{55.61}  & \num{2.60}  & \num{2.52}  & \num{6.15}  & \num{1.62}  \\
WBA   & \num{668123}  & \num{45.78}  & \num{4.87}  & \num{4.38}  & \num{10.54} & \num{2.22}  \\
WMT   & \num{886255}  & \num{118.66} & \num{6.14}  & \num{3.30}  & \num{6.82}  & \num{2.85}  \\
XOM   & \num{2057230} & \num{45.84}  & \num{9.29}  & \num{1.42}  & \num{3.43}  & \num{2.25}  \\
\bottomrule
\end{tabular}
\end{table}


\end{document}